\documentclass[manuscript]{acmart}

\AtBeginDocument{%
  \providecommand\BibTeX{{%
    \normalfont B\kern-0.5em{\scshape i\kern-0.25em b}\kern-0.8em\TeX}}}

\usepackage{hyphenat}
\usepackage{microtype}
\usepackage{color,soul}
\usepackage{afterpage}
\usepackage{rotating}
\usepackage{lscape}
\usepackage{longtable}
\usepackage{multirow, tabularx}
\usepackage{graphicx} 
\graphicspath{ {.} }

\usepackage{hyperref}
\hypersetup{
    colorlinks=true,
    linkcolor=blue,
    filecolor=magenta,      
    urlcolor=cyan,
    breaklinks=true,
}

\usepackage{pifont}
\newcommand{\cmark}{\ding{51}}%
\newcommand{\xmark}{\ding{55}}%

\begin{document}

\title{Text-based Stock Market Analysis: A Review}

\author{Kamaladdin Fataliyev}
\email{kamaladdin.fataliyev@student.uts.edu.au}
\author{Aneesh Chivukula}
\email{aneesh.chivukula@uts.edu.au}
\author{Mukesh Prasad}
\email{mukesh.prasad@uts.edu.au}
\author{Wei Liu}
\email{wei.liu@uts.edu.au}
\authornote{Corresponding author}
\affiliation{%
  \institution{University of Technology Sydney}
  \city{Sydney}
  \state{NSW}
  \country{Australia}
  \postcode{2007}
}

\renewcommand{\shortauthors}{Fataliyev et al.}

\begin{abstract}
  Stock market movements are influenced by public and private information shared through news articles, company reports, and social media discussions. Analyzing these vast  sources of data can give market participants an edge to make profit. However, the majority of the studies in the literature are based on traditional approaches that come short in analyzing unstructured, vast textual data. In this study, we provide a review on the immense amount of existing literature of text-based stock market analysis. We present input data types and cover main textual data sources and variations. Feature representation techniques are then presented. Then, we cover the analysis techniques and create a taxonomy of the main stock market forecast models.  Importantly, we  discuss representative work in each category of the taxonomy, analyzing their respective contributions. Finally, this paper shows the findings on unaddressed open problems and gives suggestions for future work. The aim of this study is to survey the main stock market analysis models, text representation techniques for financial market prediction, shortcomings of existing techniques, and propose promising directions for future research.
\end{abstract}



\maketitle

\section{Introduction}
In  the modern world, one of the keystones of the global and local economies is the stock market. Stock market is a financial market where the new issues of stocks, i.e., initial public offerings, are created and sold at the primary market whereas the succeeding buying and selling are carried out at the secondary market \cite{n-1}. Stock markets are related with huge gains but high profits also bring associated risks which can result with loss. This makes stock market prediction an appealing area but also a challenging task as it is very difficult to predict stock markets with high accuracy due to high volatility, irregularity, and noise.\par
There are two main theories which assume the movements in the financial markets are random and thus unpredictable - Efficient Market Hypothesis (EMH) and Random Walk Theory(RWH). EMH \cite{n-3} states that the price in the markets reflects the stock value accurately and responds only to new information which consists of historical prices, public information and private information. Fama categorized markets as weak, semi-strong and strong based on the plausibility of prediction \cite{n-3}. In weak markets, private and public information can be used for market forecast. In semi-strong markets, only private information can be used for market prediction as all other past information is already reflected in the price. On the other hand, strong form states that the current price includes all the past information, hence using these for prediction won't yield good results.

RWH aligns with EMH and states that financial markets are stochastic, and prices follow random walk pattern hence making it impossible to predict and outperform the market. In this way, it is consistent with EMH, especially with its semi-strong form.

In contrast to these theories, behavioral economists claim that investors can be emotional and thus their behavior can be explained using psychology based theories. Behavioral finance mainly focuses on understanding the affect of investors’ psychology in their trading strategy and on the market \cite{n-6}. \citet{n-4} shows that there is a delay between the time new information is being introduced and market correcting itself by reflecting the new information and \citet{n-5} reports that this lag is approximately 20 minutes. These results support the idea that new information can be used in stock price prediction for a short duration. Recently, a relatively new theory - the Adaptive Market Hypothesis \cite{n-327} - has been proposed to bridge EMH and behavioral finance - efficiency and inefficiency of the markets - in order to understand investor behavior better. This theory assumes that markets can be predicted by analyzing investor behavior.

\begin{figure}[ht]
  \centering
  \includegraphics[width=\linewidth]{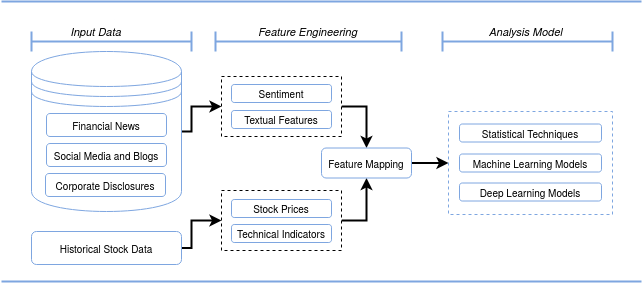}
  \caption{A standard structure of a stock market analysis workflow}
  \label{fig:process}
\end{figure}

Various approaches have been developed to analyze and beat the market. Fundamental and technical analytical techniques are mostly based on human knowledge and reasoning in areas such as locations of reversal patterns, market patterns, and trend forecasting \cite{n-90}. Although these techniques take historic stock data into consideration, most of the existing studies approach stock prices as stationary data. Also, human factors might cause inaccuracy in the results. The traditional statistical methods, including moving average, exponential smoothing and linear regression, have been used in the prediction of stock prices. However, they haven’t been really effective as these models mostly assume linear relationship in the data.

The success of machine learning methods in other time series applications made them a good prospect for stock market analysis. Support vector machines and neural networks have been vastly applied to predict price movements in the financial markets. The performance of a prediction model mostly depend on the features. Early prediction models in the field focus on the features from historical stock data. To analyze the influence of public and private information, researchers have started extracting textual data from social media, news articles and official company announcements. Figure 1 shows a standard structure of a market prediction and analysis process. Inputs from multiple sources are taken for analysis and features from these input data are fed into the model. A deep learning based representation model might differ from this structure and can include all these steps in an end-to-end model.

Market prediction is a hot research area and has become attractive to both researchers and investors. Figure 2 shows the number of publication under the topic 'Text-based Stock Market Analysis' in the last decade. As seen from the figure, the number of researches in the are has been mostly increasing. With the rise in the number of works, researchers have also surveyed the existing literature to give an overview of the area. But most of the existing survey works in the literature don’t focus on multi-modal approaches. Although they do cover some of the multi-modal works, they usually focus only on text-based works, or market analysis approaches with focus on quantitative stock data.

There aren't many surveys that cover the multi-modality and fusion side of the market analysis. \citet{n-1} is one of the rare works that focuses on fusion techniques used in the literature. The categorize the existing literature under information, feature and model fusion categories and cover stock price or trend prediction, portfolio management, risk-return forecasting and other stock market concepts. The survey focuses on the works done between 2011 and 2020. In \citet{n-311}, machine learning applications in the financial sector are covered. The authors focus on machine learning based analysis models for time series forecasting and compare them against traditional forecasting techniques. \citet{n-330} focus on text mining works for financial industry and cover FOREX rate prediction, customer relationship management and cyber security alongside stock market prediction. The authors review the works during the period 2000-2016.

Most of the review works in the literature focus on researches that only employ stock price data. \citet{n-312} gives a good review of market analysis works with focus on only quantitative stock data. The survey covers input data variables, data processing and feature representation techniques, analysis models and the performance metrics in the works. The survey focuses on the works between 2009 and 2015. \citet{n-313} is another review work that focuses on machine learning techniques in the market prediction. The authors review the works under ANN, SVM, Genetic Algorithm and hybrid techniques between 1999 and 2019.

There are also some survey papers that review text-based works only and analyze textual data sources and representation techniques. \citet{n-281} review the studies from 2007 to 2016 with a focus on web media in the form of textual data. The authors covers researches under different media types (news articles, discussion boards and social media), representation techniques for textual data and analysis models. Under representation techniques, works in the area of sentiment analysis are also surveyed. \citet{n-153} and \citet{n-314} also survey the literature for text-based works. \citet{n-153} categorizes previous studies based on the dataset used, data pre-processing techniques and machine learning models. They cover the works up to 2015. \citet{n-314} presents natural language processing techniques for market forecast. The authors analyze the works done before 2018 from three points: types of text sources employed, analysis algorithms and research results. \citet{n-316} on the other hand, puts more focus on analysis and prediction models rather than dataset and presentation techniques and covers the works done before 2019. They present a taxonomy of market analysis techniques and survey the literature based on the given taxonomy. Table 1 shows a comparative analysis of our work with with the existing review papers under various criteria.

\begin{table}
\centering
\renewcommand{\arraystretch}{1.8}
\caption{A comparative analysis of existing survey papers under various criteria:
T1 - Social media content,
T2 - News content, T3 - Official company announcements,
T4 - Traditional textual representation techniques, T5 - Deep Learning based advanced NLP techniques,
A1 - Statistical models, A2 - Machine Learning techniques, A3 - Deep Learning models}
\label{table_1}
\begin{tabular}{ cccccccccl }
\toprule
Ref & T1 & T2 & T3 & T4 & T5 & A1 & A2 & A3 & Period\\
\midrule
    \citet{n-1} & \cmark & \cmark & \cmark & \xmark & \xmark & \cmark & \cmark & \cmark & Up to 2021 \\
    \citet{n-281} & \cmark & \cmark & \cmark & \cmark & \xmark & \cmark & \cmark & \xmark & Up to 2017\\
    \citet{n-314} & \cmark & \cmark & \cmark & \cmark & \cmark & \cmark & \cmark & \cmark & Up to 2017\\
    \citet{n-316} & \cmark & \cmark & \cmark & \cmark & \xmark & \cmark & \cmark & \cmark & Up to 2019\\
    \citet{n-312} & \xmark & \xmark & \xmark & \xmark & \xmark & \cmark & \cmark & \xmark & Up to 2016\\
    Our Work & \cmark & \cmark & \cmark & \cmark & \cmark & \cmark & \cmark & \cmark & Up to 2021 \\
\bottomrule
\end{tabular}
\end{table}

With the progress in text analysis techniques, the incorporation of textual content into stock market research has become an appealing topic. As the social media, blogs and user shared news become more widespread, it is important to come up with proper techniques to analyze their influence on the market movements. This research reviews the computational stock prediction literature with the focus on text-based market analysis and make contributions in three aspects: First, we show the taxonomy of main market analysis models and review the works under these categories. Second, we present the main textual data sources and variations, clarify the main representation techniques. Finally, we show our findings and give suggestions for future work.

\begin{figure}[ht]
    \centering
    \includegraphics[width=0.5\linewidth]{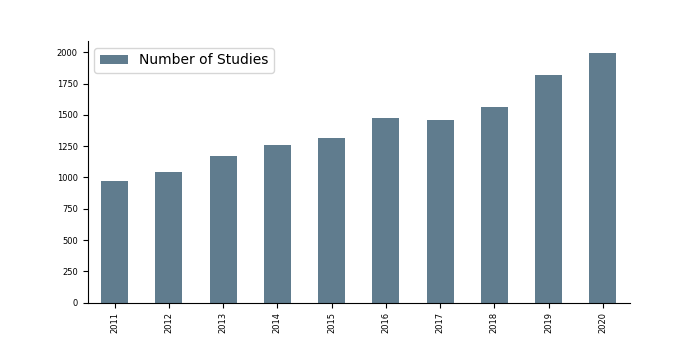}
    \caption{Yearly record counts of literature from Google Scholar by the keywords ``text-based stock market analysis'' in the last decade.}
    \label{fig:research_count}
\end{figure}

\begin{figure}[ht]
    \centering
    \includegraphics[width=\linewidth]{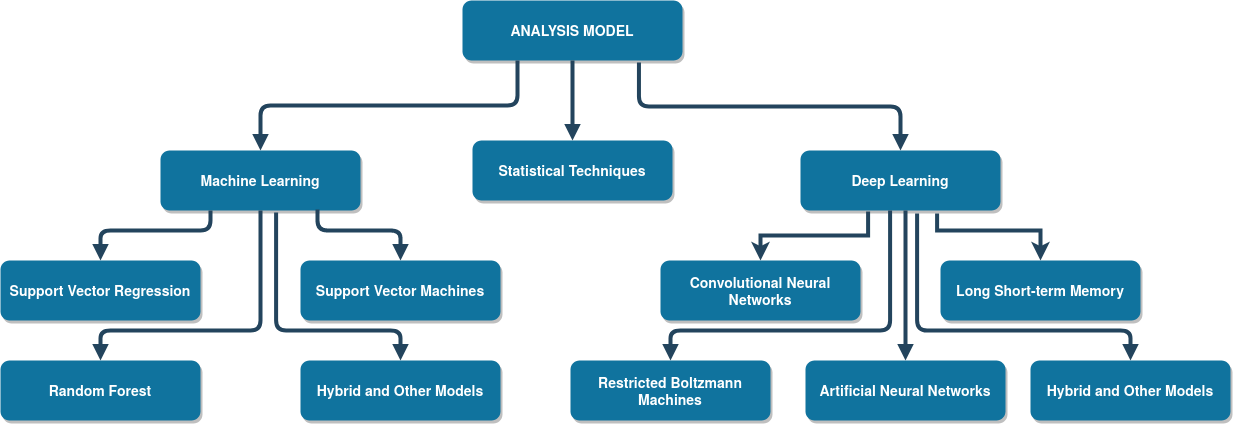}
    \caption{A taxonomy of stock market analysis models}
    \label{fig:models}
\end{figure}

The rest of the paper is structured as follows. Section \ref{sec:dt} covers the historical stock data and main textual data sources. In Section \ref{sec:fr}, main feature representation techniques including main textual representation approaches used in the area are presented. In Section \ref{sec:am}, we mention the widely used stock market analysis models and create a taxonomy.  Section \ref{sec:rw} shows the representative works in the literature and talks about their main contributions. In Section \ref{sec:con}, we make conclusions and provide suggestions for future work based on the review. 

\section{Input Data}
\label{sec:dt}
 It's been shown that using textual data alongside historical prices can improve the performance of a forecast model \cite{n-75}. \citet{n-227} show that using both technical indicators and news data yields better results than only using either technical indicators or news sentiments, in both individual stock level and sector level. \cite{n-333} show that sentiments extracted from Twitter can improve economic forecasting. \par
 In this section we cover historical stock data and textual data sources for stock market analysis.
\subsection{Financial Time Series Data}
Quantitative stock data is a time series that can consist of daily closing price, opening price, the current closing price, price change, closing bid price, volume, and closing offer price. The existing research mostly focus on stock market indexes such as DIJA, NASDAQ Index, NYSE Index,  S\&P 500,  Hong Kong Index and others. These data can be extracted for variuous frequencies such as daily, hourly, minute data etc. Studies have extracted these data from various providers including Retuers, Bloomberg, Yahoo Finance, Google Finance and others. \par
Open, close, high and low prices are the opening, closing, highest and lowest prices in a given frequency, respectively, and volume shows the total number of traded shares in that duration. Another price component that is widely used is adjusted closing price which is factors the stock's dividends, stock splits, and the new stock offerings. \citet{n-324} take open, high, low, close and trading volume for S\&P 500 and DIJA indices.  \citet{n-325} focus on high and low prices and propose an LSTM model to predict high and low prices of soybean xfutures. \par
Historical data have been used in variuos frequencies in the literature. \citet{n-189}, \citet{n-259} use daily stock data for their research. \citet{n-81} extract daily data for two stock indices – Nikkei 225 and S\&P 500 – to use in their deep learning based system. \citet{n-231} extract historical data from Google Finance API with a frequency of a quarter of hour. They derive technical indicators and use them with sentiment embeddings for market trend prediction. \citet{n-79} use financial news articles to run experiments on the tick-by-tick price data in Hong Kong Stock Exchange. \citet{n-326} categorize the stocks under nine industries: Basic Materials, Consumer Goods, Healthcare, Services, Utilities, Conglomerates, Financial, Industrial Goods and Technology. They extract the historical data for the period of two years for 88 stocks: all the 8 stocks in Conglomerates and the top 10 stocks in capital size in each of the other 8 industries. This data is used alongside twitter data for market prediction. \citet{n-298} also use the same dataset for their research. They propose a novel stock movement predictive network via incorporative attention mechanisms that uses tweets and historical price data for stock movement prediction.

\subsection{Textual Data}
\label{sec:td}
Textual data can come from various sources - social media, financial and general news, corporate announcements, blogging and micro-blogging websites. Textual content from news \cite{n-37}, financial blogs \cite{n-38}, \cite{n-39} and social media and discussion boards \cite{n-40}, \cite{n-41}, \cite{n-42}, \cite{n-43} have been used for stock market prediction.\par
In this section we review some of the works in the literature and allocate them to three categories based on the textual data: news data, social media and micro-blogging data, and data from corporate disclosures.

\subsubsection{News}
One of the main sources of the textual data is news data from websites, journals and newspapers. \citet{n-71}  and \citet{n-72} use financial news data from Reuters and Bloomberg to predict stock price movements. \citet{n-83} develop a sentiment analysis tool and used it with a financial news article prediction system called Arizona Financial Text (AZFinText). \citet{n-44} use both news data and technical features for a regression model. \citet{n-12} propose a novel model for price movement prediction based on the financial news considering casual relationships between firms. They implement context-aware text mining based on the company-specific financial news rather than general news. \citet{GEVA2014212} use standard data representations to combine market data with news text.
Although financial news is widely used as it is assumed to be less noisy, researchers have used filters in their data extraction such as all financial news, just company or product related news, stock related news and others. \citet{n-137} and \citet{n-213} claim that general financial news has limited and short-lived predictive power on future stock prices. \citet{n-132} find that fundamental information of firm-specific news articles can enrich the knowledge of investors and affect their trading activities. \citet{n-103} work on both company-specific and market related news data and use the summary of the news content rather than the whole article in the analysis. \citet{n-194} use stock related news and allocate them to five categories for their sentiment analysis model. \citet{n-74} extract titles from financial news for their S\&P 500 prediction model. \citet{n-24} employ features from news headlines for market volatility prediction. \citet{n-132} show that stock prices are sensitive to restructuring and earning issues news.

The existing literature also varies based on the source of the news data. Most of the works take the financial news from Reuters and Bloomberg \cite{n-74}, \cite{n-81}, \cite{n-24} and Yahoo! Finance \cite{n-37}, \cite{n-83}. \citet{n-111} use data from the online versions of all major Flemish newspapers to predict commodity stock prices. Hong Kong Stock Exchange focused works mostly extract financial news from a financial information platform called FINET (www.finet.hk). Tokyo Stock Exchange related studies use Nikkei as their news source \cite{n-75}, \cite{n-81}, while some Chinese works extract data from a platform called Wind, which is a widely used financial information service provider \cite{n-76}.

\subsubsection{Social Media and Micro-blogging}
The rise of the social media and micro-blogging websites created a huge source of data. Social media platforms allow users to share their opinions and emotions and enable instant interactions via posts, comments and other ways. It is not uncommon for influential people to share their opinions about political or financial situations on Twitter or other platforms and influence other market players' decisions. \cite{n-332} show that Trump's tweets carry information for short-term  market movements. Hence, these data can be really valuable in forecasting the movement in financial markets. Some of the sources include Yahoo! Finance, Sina Weibo, Google Blog, Guba and Xueqiu.

Since social media data mostly reflect personal opinions, they have widely been used for sentiment analysis applications. The research by \citet{n-40} is one of the pioneers that uses Twitter data to evaluate the effect of public mood on stock markets. \citet{n-190} and \citet{n-193} also use twit data to analyze the relationship between textual data and stock price movements, and used it for market prediction. \citet{n-55} use data from stock review blogs and employ SVM to classify the emotions and reconstructed the set using bootstrapping classifier.

Like news, the nature of social media data in the literature varies as well. Although, most of the works extract Twitter data, some of the research focus on financial social media platforms like StockTwits or discussion boards like Yahoo! Finance boards. \citet{n-11} use StockTwits data in their domain-specific sentiment analysis model with deep learning architecture. \citet{n-191} extract stock related data from Yahoo! Finance message board and show that using sentiment information from social media can help to improve the stock prediction. \citet{n-76} use a discussion board named Guba (http://www.guba.com.cn) and Xueqiu, a Twitter-like investor social network in China to extract user sentiments.

\subsubsection{Corporate Disclosures}
Corporate disclosures are considered more trustworthy and less noisy as they usually include latest official firm related information. These include important information such as quarterly earnings, legal and management information, past and current performance and challenges of the business \cite{n-152}, \cite{n-154}. The nature of these texts makes them an important and alluring data source for market forecasting \cite{n-153}. These documents, if decoded correctly, give a major insight into a company’s status, which can help to understand the future trend of the stock. Their influence on the stock returns have been analyzed in the short and long-term \cite{n-9}, \cite{n-10}, \cite{n-84}, \cite{n-157}.

\citet{n-9} analyze textual data from company ad hoc announcements to understand their effect on short-term and long-term stock index predictions. \citet{n-163} use ad hoc announcements from Thomson and Reuters to build a decision support model. They build a sentiment analysis model with focus on negation scope detection. \citet{n-159} use disclosures to get more expressive features for stock market prediction. \citet{n-82} use ad hoc announcements to build and compare multiple classifiers for stock price forecasting.\par
\citet{Balakrishnan2010OnTP} build a text classification model using data for narrative disclosures. Document-level information is processed for value-relevant information with features for risk sentiment and price momentum. Correlations are found between financial disclosure in text and financial characteristics and market performance of a firm. In terms of economic forecasts, the quality of disclosure documents and features is associated with future returns and confidence estimates of a firm. Thus predictive analytics on financial disclosure can be used to condition the modelling parameters and products life-cycles of a firm to particular industry segments and entrepreneurship ecosystems.

There are also studies that analyze reports and documents by regulatory organisations. \citet{Basu2019MeasuringIO} predict future investments by deriving industry-specific factors from SEC reports about financial performance of firms. They use Lasso and factor analysis as part of a supervised machine learning technique for driving corporate investments. 

Table 2 shows some of the research done in the textual analysis for finance area.

\begin{table*}
\renewcommand{\arraystretch}{1.8}
\caption{Stock market analysis using various types of textual content}
\label{table_2}
\begin{tabularx}{\textwidth}{@{} XXXXX@{} }
\toprule
Category &Ref & Type & Source & Market\\
\midrule
    \multirow{5}*{News Content} & \citet{n-83} & News & Yahoo Finance & S\&P 500 \\
    &\citet{n-111} & News & Flemish newspapers online & Euronext Brussels Stock Exchange \\
    &\citet{n-194} & News & LexisNexis database & Healthcare sector from S\&P 500 \\
    &\citet{n-129} & News & Reuters and Bloomberg & S\&P 500 \\
    &\citet{n-75} & News & Nikkei newspaper & Tokyo Stock Exchange \\ \hline
    \multirow{8}*{Social Media} & \citet{n-132} & Forum data (news and discussions) & Chinese stock discussion boards (sina.com, eastmoney.com) & CSI 100 \\
    &\citet{n-228} & Forum data (message boards) & Yahoo Finance & DIJA \\
    &\citet{n-109} & Forum data( message boards) & Yahoo Finance & 18 Stock from Yahoo Finance \\
    &\citet{n-76} & News and social media & News from Wind, social media from Guba and Xueqiu & CSI 100 and HKSE \\
    &\citet{n-77} & Social media based news & Sina Weibo & Shanghai-Shenzhen 300 Stock Index \\
    &\citet{n-40} & Social media & Twitter & DIJA \\
    &\citet{n-193} & Social media & Twitter & 30 companies from NYSE and NASDAQ \\
    &\citet{n-11} & Social media & StockTwits & \\ \hline
    \multirow{3}*{Official announcements} & \citet{n-161} & US corporate filings & SEC Edgar website & Corresponding stocks \\
    &\citet{n-9} & German ad hoc announcements & DGAP & DAX, CDAX, and the STOXX Europe 600 \\
    &\citet{n-163} & German ad hoc announcements & Thomson Reuters & Corresponding stocks \\
    
\bottomrule
\end{tabularx}
\end{table*}

\section{Feature Representation}
\label{sec:fr}
The performance of the prediction models mainly depends on the representation of the given data – the features. It is important to extract meaningful representations from the input data – let it be historical stock data or textual data. Hence, employing a good representation technique is a crucial task that affects the outcome of the whole model.\par
Here we cover the main representation techniques used in the literature for quantitative stock data and textual data.
\subsection{Financial Time Series Data}
In the literature, fundamental and technical analysis are widely used to derive input features for an analysis model. Technical and fundamental analysts differ on the analysis of the stock price at a given time. The major difference between these two strategies is related with the nature of market features considered by each approach \cite{n-90}. Fundamental analysts focus on the fundamental values of a stock and analyze the market based on these values. Fundamental analysis  cover financial statements, balance sheet, government policies, company market data, political and geographical circumstances for stock prediction. This kind of systematic approach allows the investors to see the changes before they are reflected in the price and thus outperform the market. Researchers have used data from market trends, financial and political news, social media platforms to understand the fundamental value of a stock in order to improve prediction accuracy.
On the other hand, technical analysts focus on historical stock data for their analysis. Financial time series data is considered noisy data and features containing meaningful information are vital for a good forecast model. Technical analysts believe that all available internal and external information are already reflected in the price and any patterns in the market data would include the fundamental values as well. Technical analysis assumes that future price is tied to some patterns in the historical data and defines technical indicators to describe these behaviours in the data. It uses historical stock data such as stock prices, trading volume and breadth as their reference point to derive mathematical indicators \cite{n-91}. Widely employed tools in the are includes charting, relative strength index, moving average, exponential moving average, moving average convergence/divergence rules, relative-strength index, on balance volumes, momentum and rate of change, directional movement indicators \cite{n-92}, \cite{n-93}.  \citet{n-33} and \citet{n-49} use historical stock data to derive a set of technical indicators to get better results in the prediction process. \citet{n-317} apply technical analysis for their deep learning based analysis model. They calculate forty-four technical indicators and use multiple feature selection approaches to find the best features for forecast. \citet{n-318} study the relationship between forecast horizon and the time frame used to calculate technical indicators. Using ten years daily price data for fifty stocks, they show that the best performance is achieved when the input window length is approximately equal to the forecast horizon. \citet{n-180} test four prediction models for two different input approaches: using technical indicators as input and representing technical parameters as trend deterministic  data. Using ten technical parameters from ten years of stock data, they show that their models perform better when they represent the technical indicators as trend deterministic data.  \citet{n-320} derive multiple technical indicators (simple moving average, weighted moving average, relative strength index and others) from ten years daily price data of ISE National 100 Index for machine learning based forecast models. \citet{n-231} employ technical indicators alongside dictionary based sentiment embeddings for market analysis. 
Statistical approach is another widely used feature engineering technique for financial time series data. Statistical methods are more about dimensionality reduction and information compression. Principal component analysis, independent component analysis, singular value decomposition are among widely used techniques. In \citet{n-33}, independent component analysis is used for dimensionality reduction and canonical correlation analysis is used to extract feature set for the forecast of next day close price. \citet{n-189} employ principal component analysis for ANN based market prediction model.
With the success of deep learning in various areas, researchers have also started exploring deep learning techniques for financial time series data. Studies either extract features beforehand and feed them into the network as input, or build an end-to-end model that can learn feature representations by itself. \citet{n-321} propose a novel end-to-end model named multi-filters neural network (MFNN) for feature extraction and price prediction on financial time series data. Using both convolutional and recurrent methods to capture different feature spaces, the model gives better performance than traditional machine learning techniques and single-structure deep learning techniques (CNN, LSTM, RNN).

\citet{n-328} use hourly price data and 25 technical indicators to predict the direction of Borsa Istanbul (BIST) 100 stocks. Their proposed end-to-end CNN-based model combines feature selection and classification steps and it outperforms other classifiers that use manually selected feature sets.

\citet{n-322} compare three RNN models for market analysis - SRNN, LSTM, GRU. The networks take open, adjusted close, high, low prices and volume as input and learn the features within the network. \citet{n-323} explore long short-term memory (LSTM), convolutional neural network (CNN) with empirical mode decomposition (EMD) and complete ensemble empirical mode decomposition (CEEMD) algorithms for one-step ahead stock market prediction. They  build three novel hybrid algorithms, i.e., CEEMD-CNN-LSTM and EMD-CNN-LSTM, to extract deep features and time sequences from stock price data.

\subsection{Textual Data}
\label{sec:tr}
In the literature, 2-g \cite{n-159}, noun phrases \cite{n-195}, sentiment words \cite{n-132}, topic modeling \cite{n-109} and bag-of-words \cite{n-194}, \cite{n-82} are widely used textual representation techniques. Bag-of-words approach is one of the most popular and basic approaches \cite{n-153} that has been employed in analyzing financial texts \citet{n-5}, \cite{n-143}. Unlike general tokenization, this method removes stop-words from the representations. Some implementations of BOW approach use stemming technique for better representation. But this approach sometimes lacks the ability to capture the semantics between words.

N-grams with various configurations have been studied for representation. \citet{n-159} compare 3-gram and 2-gram for analysis of ad hoc announcements and find that 2-gram gives better results. The research also tests 2-word combinations with feedback-based feature selection method and finds that its accuracy is better than noun phrases and 2-gram. On the other hand, \citet{n-163} employ n-gram and test unigrams, bigrams, trigrams and 4-grams for negation scope detection in their reinforcement learning model. They show that reinforcement learning with 3-grams give better predictive performance.

Another technique is called noun phrases which extracts nouns and noun phrases from a given text \cite{n-144}. An extension of this method that is called named entities that focuses on proper nouns from pre-defined categories. Semantic lexical hierarchy \cite{n-145} and semantic tagging \cite{n-146} are employed for categorization. Named entities allows for better generalization of previously unseen terms and does not possess the scalability problems associated with a semantics-only approach \cite{n-83}. Proper nouns technique extracts specific nouns and named entities without a given category set. \citet{n-37} compare these techniques for news articles and show that using proper noun method yields better results.

In the last decade deep learning methods have been successfully used for feature extraction from textual data to predict stock markets. Word embeddings is one of the techniques used to extract meaningful features from textual data. Researchers have used both domain-general and domain-specific embeddings to analyze the features. Without correct domain knowledge, it is a clear challenge to find effective matrices/measures to characterize the market movement, and such knowledge is often beyond the mind of the data miners \cite{n-307}. \citet{n-201} implement a model with CNN to train domain-specific word embeddings using StockTwits data. \citet{n-71} employ a different technique that visually interprets text-based deep learning models. \citet{n-72} use word embeddings to represent textual data. \citet{n-80} use Word2vec technique to get valuable features from text data. \citet{n-11} employ deep learning with domain-specific word embeddings to explore investor sentiment classification in financial markets. They show that including emojis lead to a better sentiment classification. The proposed model was able to detect abstract-level feature types such as sarcasm and irony. In \citet{n-81}, a deep neural generative model (DGM) with news articles using paragraph vector algorithm is used for creation of the input vector to predict the stock prices.

Sentiment analysis has been a big part of textual data usage for financial market analysis. Researchers have developed various sentiment analysis methods using text-mining and natural language processing techniques to determine the sentiment of data with respect to a specific topic. \citet{n-112} employ Naive Bayes method to classify public sentiment as positive or negative. \citet{n-193} employ TF-IDF to build a sentiment word list and a novel technique named ‘concept map’ is built to capture the relationship between sentiment words from twits about a company and its’ products. \citet{n-111} use TF-IDF to analyze news data for sentiment analysis and use sentiment polarity to assert the prediction of stock price movement. \citet{n-132} extract proper nouns to represent the event information in the news article and implement finance specific sentiment analysis (e.g. ‘bearish’ might mean something different in finance world). \citet{n-329} show that news sentiment has a significant effect on market returns. Their experiments include two different markets - equity market using DIJA and crude oil market using West Texas Intermediate crude oil. The results show that a trading system can benefit from incorporating the sentiments from relevant news into next day trading decisions.

One of the techniques for sentiment extraction is lexicon-based unsupervised learning approach. It is based on word or phrase count in a given text. \citet{n-113} use a sentiment lexicon - SentiWordNet 3.0 \cite{n-106} - to analyze the sentiments of news, and predict stock prices by training a multiple kernel learning regression model with the fusion of sentiments and prices. Another widely used version is called SenticNet 5 \cite{n-107}.

Other technique is linguistics based which involves using manually marked dictionaries. \citet{n-48} employ the Harvard IV-4 Dictionary and the Loughran–McDonald Financial Dictionary in their news based sentiment analysis model. The research shows that sentiment analysis with the given dictionaries yield better results than bag-of-words model.

Table 3 shows some of the works with their representation techniques.

\begin{table*}
\centering
\renewcommand{\arraystretch}{1.3}
\caption{Stock market analysis with textual representation techniques}
\label{table_3}
\begin{tabularx}{\textwidth}{@{} XXXXX@{} }
\toprule
Category & Ref & Content Type & Nature & Representation\\
\midrule
    \multirow{8}*{Traditional techniques}&\citet{n-37} & News content & Financial news articles &  Bag of Words, Noun Phrases, and Named Entities \\
    &\citet{n-103} & News Content & Company-specific and market related news & Bag-of-words \\
    &\citet{n-83} & News content & General financial & Proper Nouns \\
    &\citet{n-12} & News content & Company related news & Bag-of-words \\
    &\citet{n-163} & Corporate announcements & Official filings & N-gram \\
    &\citet{n-132} & Forum content & Firm-specific news & Proper Nouns \\
    &\citet{n-9} & Corporate announcements & Official filings & Bag-of-words \\
    &\citet{n-159} & Corporate announcements & Official filings & 2-gram \\ \hline
    \multirow{7}*{Deep Learning}&\citet{n-24} & News content & General Financial & Event embeddings \\
    &\citet{n-74} & News content & General financial news titles & Word2Vec \\
    &\citet{n-75} & News content & General financial & Paragraph vector \\
    &\citet{n-76} & News and social media content & General financial & Domain-specific Word2Vec \\
    &\citet{n-77} & News content & General financial & Latent Dirichlet Allocation (LDA) \\
    &\citet{n-71} & News and social media content & General financial news and stock related twits & Word embeddings \\
    &\citet{n-11} & Social media content & Finance related tweets & Word embeddings \\ \hline
    \multirow{3}*{Sentiment Analysis}&\citet{n-109} & Message board content & Stock related messages & Bag-of-words; Aspect-based sentiment \\
    &\citet{n-40} & Social media content & General tweets & Sentiments using OpinionFinder and Google Profile of Mood States \\
    &\citet{n-48} & News content & Company-specific and market related news & Bag-of-words \\
\bottomrule
\end{tabularx}
\end{table*}

\section{Analysis Model}
\label{sec:am}
In the literature, there are diverse sets of methodologies and techniques applied to stock market analysis. Statistical techniques, machine learning and deep learning algorithms have been implemented successfully. Figure 3 shows the categorization of the main techniques used in the area. Although deep learning is a subsection of machine learning, we cover it separately as these techniques have been growing in the recent years. Table 4 shows some of the studies using various statistical methods, machine learning and deep learning algorithms.\par

\begin{table*}
\renewcommand{\arraystretch}{1.5}
\caption{Stock market analysis using various analysis techniques}
\label{table_4}
\begin{tabularx}{\textwidth}{@{} XXXXX@{} }
\toprule
 Category & Ref & Market & Input Type & Model \\
\midrule
    \multirow{2}*{Statistical Techniques}
    &\citet{n-169} & S\&P 500 & Stock market data and technical indicators & GARCH \\
    &\citet{n-167} & S\&P 500 & Stock market data & ARMA \\ \hline
    
    \multirow{6}*{Machine Learning} & \citet{n-83} & S\&P 500 & Financial news and financial data & SVR \\
    &\citet{n-82} & Corresponding stocks from announcements & Corporate filings and financial data & Naive Bayes \\
    &\citet{n-231} & 20 companies from NASDAQ 100 & Financial news, technical indicators and financial data & RF, SVM, ANN \\
    &\citet{n-163} & Corresponding stocks from announcements & Ad hoc announcements and financial data & Reinforcement learning \\
    &\citet{n-111} & Brussels Stock Exchange & Technical indicators, financial news and financial data & SVM \\\hline
    
    \multirow{8}*{Deep Learning}&\citet{n-77} & CSI 300 & Financial news and financial data & RNN with GRU \\
    &\citet{n-10} & CDAX & Official announcements and financial data & LSTM \\
    &\citet{n-189} &  S\&P 500 Index EFT & Financial data and technical indicators & ANN \\ 
    &\citet{n-24} & S\&P 500 & Financial news, social media and financial data & CNN \\
    &\citet{n-74} & S\&P 500 & Financial news and financial data & RCNN (RNN + CNN) \\
    &\citet{n-34} & Chinese Stock Market & Financial news and financial data & Hybrid Attention Network with self-paced learning mechanism \\
    &\citet{n-80} & Taiwan Stock Exchange & Financial news, financial data and technical indicators & CNN + LSTM \\
    &\citet{n-196} & IBB Index & Financial data & Autoencoder \\ \\
\bottomrule
\end{tabularx}
\end{table*}

\subsection{Statistical Techniques}
Time series in stock market analysis is a chronological collection of observations such as daily sales totals and prices of stocks. Linear regression, auto-regressive moving average (ARMA), auto-regression integrated moving average (ARIMA), generalized autoregressive conditional heteroskedasticity (GARCH) and the smooth transition autoregressive (STAR) have been widely used in time series analysis.\par
ARMA employs auto-regressive models to understand the momentum and mean aversion effects, while ARIMA was implemented to analyze the non-stationarity in the time series data and attempt to reduce it to stationary series. Researchers have applied GARCH \cite{n-18} and STAR \cite{n-19} as more advanced forms of statistical techniques. \citet{n-18} use GARCH to focus on additive outliers (AO) – corrected returns. The research shows that the one-step ahead forecasts of volatility based on AO-corrected returns outperform the forecasts from GARCH and GARCH-t models for unadjusted returns. ARMA model is employed by \citet{n-167} for S\&P 500 and London Stock Exchange stocks forecasting.\par
There are also some hybrid models that uses statistical techniques. \citet{n-168} propose a hybrid model by combining ARIMA with support vector machines and apply it for multiple stock prediction. \citet{n-232} integrate artificial neural networks with ARIMA for British pound/US dollar exchange rate prediction.\par
The main shortcoming of the statistical techniques is assuming linearity in the stock data and ignoring the stochasticity of stock markets.
\subsection{Machine Learning}
Although Deep Learning is a subfield of Machine Learning, we put Deep Learning in Section \ref{sec:dl} as an independent subsection due to its significant research advancements in the past decade.

Financial markets have a non-stationary and non-linear nature and are considered dynamic, chaotic and noisy \cite{n-235}. Machine learning algorithms are able to approximate non-linear functions and find underlying patterns which makes them useful in forecasting stock movements. On the other hand, these algorithms are prone to over-fitting. They often have difficulty finding the global optimum and can easily fall into local minima \cite{n-237}.

Various machine learning models have been tried for stock market analysis: support vector machines(SVM) \cite{n-20}, support vector regression (SVR) \cite{n-33}, \cite{n-331}, k-nearest neighbors (KNN) \cite{n-233}, decision tree classifiers \cite{n-234}, random forest (RF) \cite{n-278}, fuzzy system \cite{n-21} and hybrid methods \cite{n-22}, \cite{n-23}. Below, we cover some of these techniques used for financial market analysis.

\subsubsection{Support Vector Machines}
SVM is a supervised model that tries to minimize the upper threshold of the error of its classifications \cite{n-125} by transforming the training samples into a greater space. The transformation is done with the help of kernel functions. It is a classification technique that maps training samples into points in a space and tries to create a gap between them, so new examples can be classified accordingly.\par
With ANN, SVMs are the most used machine learning techniques in stock market prediction  \cite{n-170}. \citet{n-37} employ an SVM-based model for market prediction using financial news and stock prices. \citet{n-259} build three models – ANN, ANFIS and SVM - to predict daily stock price movements of Borsa Istanbul BIST 100 Index. In the experiments, SVM outperforms the other two models. \citet{n-261} employ SVM with feature selection techniques for price index forecast. \citet{n-48} implement SVM based prediction model that incorporates financial news. \citet{n-282} experiment with adaptive parameters by incorporating the nonstationarity of financial time series into SVM. The results show that the SVM with adaptive parameters outperforms the standard SVM in financial forecasting.\par
\citet{n-283} propose an evolving least squares support vector machine (LSSVM) learning paradigm with a mixed kernel to predict the movement direction in financial markets. The model integrates a genetic algorithm (GA) for feature selection, and another GA for parameter optimization of the model.

\subsubsection{Support Vector Regression}
SVR operates similar to SVM, with the main difference of being a regression technique rather than a classification one \cite{n-147}. This technique attempts to transform training samples into multi-dimensional hyperplane in order to minimize the fitting error and maximize the goal function. Like SVM, SVR is also widely used for financial applications. \citet{n-33} employ an SVR-based model for one-day-ahead prediction of closing prices.\par
An SVR model with Self-Organizing Feature Map (SOFM) is used by \citet{n-185} for the analysis of Taiwan Stock Market Index (FITX). \citet{n-23} propose to integrate SVR in the first stage and fused SVR, ANN, and random forest (RF) in the second stage that combined SVR-ANN, SVR-RF, and SVR-SVR models for prediction. 
\subsubsection{Random Forest}
Random forest (RF) \cite{n-177} is an ensemble learning algorithm that uses decision trees as the base learner. It integrates the tree bagging approach \cite{n-275} with random subspace technique. As the assumption of prior distribution is not necessary in RF, the technique has been well implemented in financial applications \cite{n-178}. \citet{n-180} employ ANN, SVM, RF and NB models for prediction of CNX Nifty and S\&P Bombay Stock Exchange (BSE) Sensex price indices. Their experiments show that RF algorithm outperforms SVM in the market prediction.\par
\citet{n-150} compares three ensemble models – RF, AdaBoost, kernel factory – against single classifier models such as NN, SVM, kNN, and logistic regression. They test the models on the data from 5767 publicly listed European companies for one year ahead prediction. The results show that RF outperforms all the other algorithms. \citet{n-278} employ RF regression for short-term price prediction using online data sources. Besides RF, they also test NN, SVR and boosted regression tree. Their results indicate the boosted regression tree (BRT) and the Random Forest Ensemble (RFR) as the best models for predicting the 1-day ahead stock price.
\subsection{Deep Learning}
\label{sec:dl}
Deep learning is a particular kind of machine learning that achieves great power and flexibility by learning to represent the world as a nested hierarchy of concepts, with each concept defined in relation to simpler concepts, and more abstract representations computed in terms of less abstract ones \cite{n-56}. The key advantage of deep learning is the capability of representation learning and the semantic composition empowered by both the vector representation and neural processing \cite{n-306}. Deep nonlinear topology in neural networks can successfully model complex real-world data by extracting robust features that capture the relevant information \cite{n-240} and achieve even better performance than traditional machine learning methods \cite{n-241}.

The rising success of deep learning models in the fields such as pattern recognition, image and speech processing have made them a good alternative for financial time series analysis. Artificial neural networks(ANN) \cite{n-189}, convolutional neural networks (CNN) \cite{n-206}, deep belief networks (DBN) \cite{n-242}, recurrent neural networks (RNN) \cite{n-62},  stacked autoencoders \cite{n-243}, a new model of RNN – Long Short-term Memory (LSTM) \cite{n-209} have successfully been applied to different real-world applications. \citet{n-62} implement RNNs for feature learning and apply reinforcement learning module for trading decision making using deep representations.

Below, we cover some of the main deep learning architectures used in the area of stock market prediction.\par

\subsubsection{Artificial Neural Networks}
ANNs try to imitate the learning process of humans and work on identifying patterns in the data. The basic unit is called a neuron which receives input and generates an output based on the given function. Interconnected units are attributed with weights and the model attempts to minimize the error by optimizing parameters.

The study by \citet{n-236} was the first to apply ANN models in the financial domain.  \citet{n-89} review the neural network applications in financial markets. \citet{n-188} implement a neural network model that takes stock prices and market index as input for Sao Paulo stock exchange (BOVESPA). An automatic trading system was proposed using ANNs by \citet{n-86}. \citet{n-49} use adaptive neuro-fuzzy inference systems (ANFIS) model for prediction of stock close prices. \citet{n-50} implement a Bayesian-regularized feed-forward ANN model for the one-day-ahead market trend forecasting. \citet{n-189} use ANN with principal component analysis (PCA) for the prediction of S\&P 500 ETF (SPY) returns.
\subsubsection{Convolutional Neural Networks (CNN)}
CNN is a feed-forward neural network that contains more hidden layers than a standard neural network model. A typical CNN architecture includes convolutional layer, pooling layer and fully connected layers. The convolution, pooling and dropout operations makes it possible to have much deeper architectures without over-fitting the data. Compared to some of other deep learning models. CNNs require more data. \par
After their success in areas such as image processing, CNNs have been employed for stock market analysis as well. \citet{n-69} use their success in image processing and build a 2-D deep CNN for trend forecasting. The research transforms the input data into 2-D image for their CNN-based model. They  use 15 different technical indicators each with different parameter selections are utilized alongside with stock prices. Then each indicator creates data for 15-day period and 15x15 sized 2-D images are created. Their algorithmic trading model is used for stock prices of Dow 30 stocks and daily prices of nine Exchange-Traded Fund (ETFs). \citet{n-68} also apply CNN to stock price movement forecast using ETFs. \citet{n-305} build a deep convolutional fuzzy systems (DCFS) and fast training algorithms for the DCFS for the forecast of Hang Seng Index (HSI) of the Hong Kong stock market.
\subsubsection{Long Short-term Memory (LSTM)}
RNN can process raw text in sequential order to learn context specific features. On the other hand, vanishing gradient problem and short context dependencies affect their real-world performance \cite{n-245}. A new model of RNN – LSTM \cite{n-209} has been proposed to overcome that issue. LSTM employs memory cells to process inputs with long dependencies \cite{n-209}. It uses hierarchical structures and a large number of hidden layers for feature engineering. LSTM has widely been used in financial and other time series models. \citet{n-247} implement an LSTM based model that takes event information, backlog and stock prices as input for prediction of a single stock price. In \citet{n-59}, CNN and LSTM model structures are implemented together where CNN is used for stock selection and LSTM is used for price prediction. \citet{n-63} employ LSTM model for the prediction of a single stock market index. Their experiments show that LSTM outperform memory-free classification techniques such as random forest and logistic regression classifier. \citet{n-75} use paragraph vector and LSTM for their model with textual data. Their model is tested on data from fifty companies listed in Tokyo Stock Exchange. \citet{n-98} use technical indicators and stock data as the input for their LSTM based forecast model and test it on IBovespa index from the BM\&F Bovespa stock exchange. They show that LSTM is able to learn even with input with very large dimension.
\subsubsection{Restricted Boltzmann Machines (RBM)}
RBM is a different type of ANN model that can learn the probability distribution of the input set \cite{n-251}. RBM consists of visible and hidden layers and the units in the layers are not connected. RBM learning process is performed multiple times on the network \cite{n-251}. \par
RBM has been successfully applied to financial market forecasting. \citet{n-244} develop an improved deep belief network with continuous restricted Boltzmann machines to exchange rate prediction. The model is tested with British pound/US dollar (GBP/USD), Indian rupee/US dollar (INR/USD), and Brazilian real/US dollar (BRL/USD) weekly exchange rates data sets. Their results show that the improved model works better than conventional neural networks such as feed forward nets.

\citet{n-237} propose a 3-layer deep belief networks for stock market prediction. Their DBN involves two restricted RBMs to capture the feature of input space of time series data. They employ particle swarm optimization algorithm to optimize the RBMs structure. \citet{n-248} build an RBM based model for short-term stock market trend prediction.

\section{Representative Works and Contributions}
\label{sec:rw}
It has been shown that public mood and emotions have influence on trading strategies, which affects the prices \citet{n-118}. Research in behavioral finance also shows that the stock market movements are affected by emotion \cite{n-130} , financial news and social media content \cite{n-83}, \citet{n-132}. \citet{n-135} experiment with multiple textual data sources and show that the impact of different types of social media varies significantly.

In this section we review the literature and focus on the main works and their contributions in terms of the textual data they use, the representation types for the data and analysis models. We review some of the experiments and represent their results and comparison. Table 5 shows some of the representative works in the are.

\subsection{Textual Data}
While working with social media data or official company disclosures, researchers usually take the whole content of the textual data. But there are various approaches to working with news articles such as taking just the news title, the whole article content, or the summary of the article. \citet{n-71} compare using news title and content as input and conclude that their proposed deep learning model does not benefit from using additional content; so they focus on working with news titles. \citet{n-103} focus on the summaries of the articles and compare the techniques using news article summarization and full-length articles. They use both company-specific and market-related news articles and run experiments on five year data from Hong Kong Stock Exchange at at the individual stock, sector index, and market index levels. Their results show that using news article summarization for the prediction improves the performance and is better than using full-length articles.

In the literature, studies have extracted different kind of textual data for market analysis \cite{n-103}: company related textual data, both company and market related data, product data, sector related and texts about related companies have been analyzed for prediction tasks. \citet{n-48} extract both company-specific and market related financial news written in English from a major financial news vendor called Finet. \citet{n-310} extract financial and economic news articles from Naver. \citet{n-284} use financial company-specific news articles alongside with technical indicators to build trading strategies.  Their results indicate that combining news articles with technical indicators can lead to higher returns. \citet{n-132} use firm-specific financial news articles alongside with data from discussion boards to analyze their effects on investors’ trading activities. \citet{n-193} extract social media data from Twitter mentioning pre-selected 30 companies directly or indirectly (i.e. their products or services.). For example, for the company Apple, their lookup keywords include ‘AAPL’ (market code of the company), ‘iPad’, ‘iPhone’ (company products), etc.. \citet{n-135} analyze data for a specific firm on a daily basis for a large range of firms rather than focusing on one company or sector. The research includes data from overall media instead of a specific business domain which can lead to high misclassification rate and spurious correlations.

Some of the works in the existing literature analyze the data based on the categories within Global Industry Classification Standard (GICS). citet{n-12} focus on asymmetric relationship of firms within GICS sector. They analyze company-specific financial news considering casual relationships between firms (i.e target firm and casual firms).  The proposed method is tested on Korean market dataset and outperforms traditional algorithms that assumes there is bidirectional influence between all firms. The research shows that, even if there is no firm related news, the model can use news related to the causal companies to predict the target firm’s price directional movement. Another research is done on the stocks from the Health Care sector in GICS. \citet{n-194} take news articles from LexisNexis database and allocate them to different categories according to their relevance to the target stock, its sub-industry, industry, group industry and sector (according to GICS as in \citet{n-195}.  The results show that using multiple news categories with the Multiple Kernel Learning outperforms SVM and kNN models using single news category. Another finding is that using lower numbers of news categories reduces the model performance. On the other hand, \citet{n-310} claim that GICS is limited in finding relevance regarding stock prediction and propose a model that incorporates heterogeneity and searches for homogeneous groups of companies which have high relevance. The research combines data from the target company and its homogeneous cluster (developed using k-means clustering) and the tests show that the proposed model outperforms the GICS system based and individual company level forecast systems.

Textual data from multiple sources are sometimes analyzed to improve prediction performance. \citet{n-71} use financial news data from Reuters and Bloomberg and stock-related social media data from Twitter for their deep learning based end-to-end model. \citet{n-135} analyze the influence of social and conventional media on short-term stock market returns and risks. They show that when used separately data from social media gives better performance, but sentiment analysis based on both social and conventional data may increase the accuracy. \citet{n-11} focus on social media data from finances-specific platform called StockTwits. The authors also include emojis in their data to explore their effect on investor sentiment analysis and show that it significantly improves sentiment classification in traditional algorithms. \citet{n-303} build a multi-source heterogeneous data analysis (MHDA) price prediction model by combining stock data, news event data and investor comments from financial discussion boards. The model is tested on the data of palm oil features. A future specific news events analysis module is developed to analyze palm oil related news articles.

\subsection{Textual Representation}
In the literature, bag-of-words, n-grams, named entities are widely used as textual representation techniques. \citet{n-194} use bag-of-words method to represent textual data. \citet{n-84} employ bag-of-words technique with TF-IDF and represent textual data from corporate disclosures as feature vectors. They show that using textual data with stock data improves prediction performance. \citet{n-292}use bag-of-words to build a term vector from proper nouns and sentiment words. They propose a tensor-based stock information analyzer named TeSIA for market forecast.

With the success of deep learning algorithms, researchers have started employing these techniques for feature representation and engineering. \citet{n-291} build a tensor-based event-driven LSTM model for market movement forecasting. The authors use tensors to preserve the interconnections between different kind of features – technical indicators and media sentiment.  The proposed model shows superority when compared with  state-of-the-art algorithms, including AZFinText, eMAQT, and TeSIA. \citet{n-79} propose a model using deep learning representation architecture for feature representations and a supervised learning algorithm—extreme learning machine—as the prediction model. The authors show that using deep learned feature representation together with extreme learning machine can improve the prediction accuracy.

Word embeddings method has been well implemented recently. Word2vec and Glove are two widely used word embedding algorithms \cite{n-11}. \citet{n-74} employ Word2Vec model to represent financial news titles from Reuters. \citet{n-76} use Word2Vec to extract domain-specific word embeddings from Chinese financial news. The news corpus data is used to extract stock-related events. In \citet{n-298} word embeddings technique is used to represent the textual data, which are fed into a bidirectional gated recurrent unit (Bi-GRU) network and obtain the tweet-level contextual embeddings. The authors propose a novel stock movement predictive network (SMPN) via incorporative attention mechanisms using Twitter data and historical stock data. The incorporative attention combines local and contextual attention mechanisms to clean the contextual embeddings by using local semantics which reduces the noise in the features and improves the model performance.

The efficiency of domain-specific and domain-general word embeddings have also been explored in the literature. \citet{n-11} use GloVe and Word2Vec to extract domain-specific word similarities from StockTwits social media data. The research shows that domain-specific word embeddings capture the investor sentiment better than domain-general embeddings. \citet{n-201} use CNN to build a large scale sentiment lexicon for stock market. The authors train domain-specific word embeddings using StockTwits data and show that domain-specific sentiment-oriented embeddings outperform domain-general word embeddings such as Word2Vec model. \citet{n-10} employ representation learning approach (i. e. pre-training word embeddings) by using a different, but related, corpus with financial language and then transfer the resulting word embeddings to their dataset from German ad-hoc announcements in English from DGAP. The authors build a deep learning model with LSTM and compare it with traditional machine learning algorithms using bag-of-words approach. The results show that the LSTM model are superior, especially when the authors further pre-train word embeddings with transfer learning.

Paragraph vectors \cite{n-75} and document embeddings \cite{ChenASA} have been implemented to analyze textual data. \citet{n-75} employ Paragraph Vector approach to capture the distributed representations of news articles from the morning edition of the Nikkei newspaper. The authors show that the distributed representations technqiue outperforms bag-of-words based methods. \citet{ChenASA} propose financial indices to classify firm characteristics by building document embeddings from pre-trained language models on the unstructured corpus about business operations and financial performance in the annual SEC filings of a firm. Compared to traditional approaches for word embeddings, the proposed document embeddings are context-sensitive, semantically meaningful and able to model complex dependencies in text.

Some of the works focus on the event representation and employ various techniques to extract event information from textual data. \citet{n-284} extract events from Reuters news articles for the companies in the FTSE350 stock index. The research employs the ViewerPro tool for event extraction which relies on domain-specific knowledge. \citet{n-24} develop a neural tensor network to learn event embeddings from financial news titles. The research represents events using dense vectors and analyze the combined influence of long-term events and short-term events on stock price movements.
 
Sentiment analysis has been widely studied for market analysis. \citet{n-48} incorporate Harvard IV-4 Dictionary and Loughran–McDonald Financial Dictionary sentiment dictionaries to create a sentiment space for their news-based model. They show that the sentiment analysis model outperforms bag-of-words model at the individual stock, sector and index levels, but models using just the sentiment polarity don’t perform well. \citet{n-231} use McDonald dictionary and AffectiveSpace 2 \cite{n-46} to extract sentiment emebeddings from summaries of financial news articles  for twenty most capitalized companies listed in the NASDAQ 100 index. 
\citet{n-115} propose sentimental transfer learning approach that uses sentiments extracted from news-rich stocks (source stocks) to improve the prediction performance of the news-poor stocks (target stocks). The authors extract the financial news data from from Finet that includes both company-specific and market related news. The sentiment information is extracted using three dictionaries: Loughran-McDonald, Harvard IV-4, and SenticNet 3.0. The authors test the proposed technique on the data of Hong Kong Stock Exchange stocks from 2003 to 2008 and the results show that sentiment transfer learning can improve the prediction performance of the target stocks. \citet{n-304} apply Word2Vec model for textual representation of user comments and employ CNN based classifier to extract users’ bullish-bearish tendencies. The experiments using data from Shanghai and Shenzhen 300 constituent stocks show that  users’ bearish tendencies are reflected in stronger market volatility and higher market returns, and the consistency of online users’ tendencies has a positive impact on market volatility. \citet{n-40} employ two sentiment analysis tools to evaluate the daily Twitter feeds for sentiment extraction: OpinionFinder to get positive vs. negative sentiment values, and Google-Profile of Mood States (GPOMS) for more detailed analysis that classified the mood into 6 dimensions of sentiment (calm, alert, sure, viral, kind, and happy).

\afterpage{%
\clearpage
\thispagestyle{empty}%
\begin{landscape}
\renewcommand{\arraystretch}{1.3}
\begin{longtable}{ p{.07\textwidth} p{.10\textwidth} p{.10\textwidth} p{.13\textwidth} p{.10\textwidth} p{.15\textwidth} p{.35\textwidth}}

\caption{Representative works in stock market analysis with text data}
\label{table_5} \\
Category & Ref & Text Type & Representation & Market & Model & Notes \\ [0.5ex] 
 \hline\hline
\multirow{3}*{Data} & \citet{n-48} & News & Bag of Words & HKSE & SVM & This research uses both company-specific and market-related news. It also builds a sentiment space using 2 well known dictionaries.\\
&\citet{n-103} & News & Bag of Words & HKSE & SVM & This research takes both company-specific and market related news. It extracts the summary of the text rather than using the whole article \\
&\citet{n-194} & News & Bag-of-words & S\&P 500 & Multiple kernel learning & It shows that using multiple news categories helps with the forecast performance \\
\hline
\multirow{3}*{Feature}& \citet{n-75} & News & Paragraph vector & Tokyo Stock Exchange & LSTM & It shows that distributed representations of textual information are better than the numerical-data-only methods and bag-of-words based methods \\
&\citet{n-37} & News & Bag of Words, Noun Phrases, and Named Entities & S\&P 500 & SVM & The research experiments with various representation techniques and compares them. \\
&\citet{n-54} & Social Media &  & Shanghai Stock Exchange & Deep Random Subspace Ensembles & This research proposes a novel framework that uses wisdom of crowds with technical indicators. It also extends a Chinese sentiment dictionary to make it more finance-specific. \\ [1ex]
\hline
\end{longtable}
\end{landscape}
\clearpage
}

\afterpage{%
\clearpage
\thispagestyle{empty}%
\begin{landscape}
\renewcommand{\arraystretch}{1.3}
\begin{longtable}{ p{.07\textwidth} p{.10\textwidth} p{.10\textwidth} p{.13\textwidth} p{.10\textwidth} p{.15\textwidth} p{.35\textwidth}}

\caption{Representative works in stock market analysis with text data(contiuned)}
\label{table_6} \\
Category & Ref & Text Type & Representation & Market & Model & Notes \\ [0.5ex] 
 \hline\hline
\multirow{3}*{Feature}& \citet{n-132} & Forum & Proper nouns & CSI 100 & SVR & This research uses domain specific sentiment analysis with firm-specific news data \\
&\citet{n-83} & News & Proper nouns & S\&P 500 & SVR &  It builds a sentiment analysis system and shows  traders’ contrarian character – buy after bad news and sell after good news. \\
&\citet{n-163} & Official announcements & n-gram & corresponding stocks from the text data & Reinforcement Learning & This research shows that negations affect investor’s perception of the stock market and reinforcement learning can make the negation scope detection more accurate. \\
\hline
\multirow{3}*{Model} &\citet{n-74} & News & Word2Vec & S\&P 500 & RCNN &  This research proposes a novel RCNN model by combining RNN and CNN for different tasks \\
&\citet{n-76} & News and Social Media & Word2Vec & CSI 100 and
HKSE & A novel tensor-based computational model & This research develops a novel tensor-based prediction model. It uses domain specific word embeddings. \\
& \citet{n-71} & News and Social Media & Word Embeddings & S\&P 500 & Deep NN & It builds a visually interpreted end-to-end framework for end users \\ [1ex]
\hline
\end{longtable}
\end{landscape}
\clearpage
}

Studies also vary based on the level they focus for sentiment extraction. \citet{n-135} extract sentence-level and document-level sentiment polarity from social media – blogs, forums and Twitter, and conventional media – newspapers, television and business magazines using Naive Bayes. \citet{n-109} propose a new feature called ‘topic sentiment’ for market forecast. They extract topic sentiment which represents the sentiments of the specific topics of the company (product, service, dividend and so on), from Yahoo Finance Message Board using Latent Dirichlet Allocation (LDA). 

\subsection{Analysis Model}
Financial Technology (FinTech) is producing innovations in economics and finance (EcoFin). FinTech consists of AI areas such as logic, planning, reasoning, modeling, simulation in expert systems, decision support systems and natural language processing. It is undergoing rapid development in problem solving,self-correcting and decision-making with the introduction of machine learning areas such as signal processing, pattern recognition, mathematical modeling, data analytics, knowledge discovery, deep learning, representation learning, computational intelligence, complex systems and optimization methods. Machine Learning for FinTech has to deal with multisource, multimodal and heterogeneous data for high-dimensional, sequential and evolving economic-financial models \cite{9090124}.\par
As stated in Section \ref{sec:am}, statistical and machine learning methods are widely used for stock market prediction. ARIMA, SVM, ANN, RF and hybrid models have been implemented to incorporate textual data and stock data features. \citet{n-84} creates a hybrid model by combining ARIMA with SVR. After getting feature vectors from textual data,  ARIMA is used to analyze the linear part. Finally, an SVR model based on textual feature vector is developed for the nonlinear part. The authors compare the proposed model with a pure ARIMA model and a hybrid ARIMA and SVM model that uses stock data only. The results show that the proposed model outperforms the baseline techniques for the prediction of six pre-selected stocks. \citet{n-163} employ reinforcement learning to build a tailored trading decision support model with a focus on negation scope detection. The results also show that negations affect investor’s perception of the stock market and reinforcement learning can make the negation scope detection more accurate. \citet{n-190} compare SVM with kNN and Naive Bayes based sentiment analysis models. They use company-specific social media data and develop a novel sentiment analysis model for stream-based active learning. They implement an active learning model using SVM classifier to analyze the relationship between company related twit data and their stock price. The results show that Naive Bayes under-performed compared to SVM and using kNN was not computationally efficient – the model was too slow.\par
Researchers have also started using deep learning techniques for market analysis and forecast. \citet{n-75} build a news-based analysis model using LSTM to predict next day’s prices and compare their model with SVM, MLP and RNN.  In their trading simulations, LSTM outperforms the other models. \citet{n-54} propose a novel model called Deep Random Subspace Ensembles (DRSE) that uses public sentiment and technical indicators for market forecasting. The proposed DRSE model combines deep learning algorithms and ensemble learning.  They take Tsinghua Sentiment Dictionary 1.0 for sentiment analysis but extend it to enable more finance based word analysis. \citet{n-71} builds a deep learning based system called DeepClue for end-users - stock traders from public/private funds. The proposed deep regression model consists of four layers: a word representation layer, a bigram representation layer, a title representation layer, and a feed-forward regression layer. DeepClue bridges the analysis model and end-users by visually interpreting the main points of the prediction model. \citet{n-81} develop a novel Deep Neural Generative model for the forecast of daily stock price movements using financial news. They employ paragraph vector technique to represent textual data, and the comparison against support vector machines and multi-layer perceptrons and show that the deep learning based model outperform the traditional techniques in the forecast of both mentioned markets. \citet{n-227} use LSTM network for analysis model and compare it with SVM and MKL based models. The authors  use sentiment analysis to represent textual data to use for market prediction alongside with technical indicators from stock data. The proposed model outperforms the baseline models in terms of prediction accuracy when using both information sources.\par
Hybrid deep learning techniques have been proposed to improve the model performances.  \citet{n-74} build a hybrid model called RCNN by combining RNN and CNN to make use of both models’ advantages. The experiments show that the proposed model is better than a CNN model and using textual data and technical indicators as input data has positive influence on the model performance. Another hybrid model called RNN-boost is used to predict the stock volatility \cite{n-77}. LDA features and sentiment features are extracted from social media data to be used with technical indicators. The proposed model incorporates RNN and Adaboost and achieves an average accuracy of 66.54\% and a best accuracy of 70.17\%. The RNN model uses Gated Recurrent Units (GRU) to predict the stock price.

\section{Conclusion and Suggestions for Future Work}
\label{sec:con}
Stock market prediction has been an attractive area due to its potential influence to investment income. But financial markets carry noisy data and have non-linear nature. More advanced prediction methodologies have been developed to incorporate textual data for price forecast. The research covering the area of stock market analysis with textual data mostly focus on three main aspects, textual representation techniques, analysis models, and textual content.
After the review of the literature, we can make the following conclusions:
\begin{itemize}
\item Even after the significant advancements of deep learning based methods, they haven’t been widely implemented in finance: conventional machine learning and statistical methods are main approaches in the area.
\item Most of existing methods use textual data from a single source, despite of the availability of multiple data sources, for stock market analysis.
\item Simplistic text representations such as bag-of-words representation techniques are still widely used for textual data.
\item Hybrid analysis models have shown great potentials and can usually lead to better results than conventional methods.
\end{itemize}

Extensive research has been done to understand the influence of textual content on the stock market movements. With the huge increase in the amount of available data, the task has become more important and more challenging. The technological progress – increase in the computing power, success of more advanced machine and deep learning techniques, progress in natural language processing algorithms has enabled researchers to analyze vast textual data. As stock markets carry huge importance in the economy, it is necessary to implement the latest methodologies in the area. 

Based on our survey, we propose the following three areas (directions) for future research and advancement.

\subsection*{Future Direction 1: Textual Representation}
In the literature, most of the existing works implement bag-of-words approach for representation. Named entities, proper nouns have also been used to get more advanced features from textual data. But due to the simplistic representation, important textual features can be skipped in the analysis. The key problems that previous methods, such as bag-of-words, failed to address are the absence of word ordering, lack of context and the curse of dimensionality: a new sentence on which the model is tested is likely to differ from all the word sequences that were seen while training with a resulting data-sparsity problem due to the increasing number of unique words, the vocabulary size and thus the representation size for each word or document\cite{n-309}.

The success of more advanced machine learning and deep learning techniques in capturing the underlying patterns in given data makes them appealing for the task. It has been shown that CNNs can learn using character level representation with no prior knowledge \cite{n-279}. Researchers have started employing deep learning based Word2Vec representation techniques to analyze textual content for market forecast \cite{n-74}, \cite{n-76}. Deep learning based textual analysis and natural language processing methodologies can be employed to understand the financial texts better.

There are also studies like  \citet{n-68}, \citet{n-69} that represent technical indicators and price data using 2-D image format for their CNN-based analysis model. These techniques can be employed and extended to incorporate textual content from social media and financial news to build a more comprehensive feature map. Another implementation of deep learning is done by \citet{n-280}. The proposed novel Stock2Vec technique can be further explored and extended.

Another aspect is using more advanced and powerful representations than word embeddings. \citet{n-74} show that sentence embedding representation can outperform word embeddings. \citet{n-24} implement event embeddings for text analysis in market prediction. But these cases are rare and sentence and event embeddings haven’t been widely explored for market analysis.

Also, as sentiment analysis is widely used to understand the influence of investor sentiment and public mood on price fluctuations and regular movements, it is necessary to improve the sentiment dictionaries for financial corpus. Most existing work has relied on general-domain sentiment analyses \cite{n-281} which can miss finances-specific meanings of the words. Therefore, it is crucial to work on extending dictionaries with finance-specific corpus.

\subsection*{Future Direction 2: Learning Model}
As seen from the survey, the area is still dominated by traditional machine learning based models. Although these models can understand non-linear relationships, they are prone to over-fitting and heavily depend on data pre-processing and representation. Therefore, it is important for researchers to build more deep learning based approaches to achieve better results.

Representation learning with transfer learning is another direction that can be explored. \citet{n-10} show that this approach can outperform traditional machine learning techniques using representations from different but related corpus and can outperform traditional machine learning models. Also, more recent deep learning techniques such as generative adversarial network (GAN) models, graph deep learning models can be studied. The applications of graph-based deep learning not only enables mining the rich value underlying the existing graph data but also helps to naturally model relational data as graphs \cite{n-315}.

Adversarial machine learning is a paradigm to design and analyze machine learning algorithms in the presence of an active adversary. \citet{ijcai2019-810} demonstrate the use of adversarial training in prediction of stock market movements assumed to be a classification problem with adversarial perturbations to simulate the stochasticity in stock price. \citet{45839} apply adversarial perturbations to word embeddings in text useful for semi-supervised learning.

As shown by \citet{n-74} and \citet{n-77}, creating hybrid deep learning models and using strong sides of each model where needed, can lead to better prediction results. Even with traditional machine learning techniques, hybrid models mostly outperform pure, single models as they can capture different kinds of patterns in the data. So, hybrid models integrating various machine learning and deep learning models can be studied and implemented.

\subsection*{Future Direction 3: Textual Content}
Most of the work in the stock market prediction area incorporate a single source of textual data: financial news, social media and blogs or corporate announcements. These data sources differ in the way they affect the financial markets. Public mood from social media, news data and author sentiment from financial news, the officiality of ad hoc announcements can influence the prices in different ways. Therefore, it is important to also look into integrating data from various source to understand the effect of textual data on markets.

As corporate announcements are official reports and filings, they carry  big importance and their analysis can help in the prediction process. But in the area, this source has been underused compared to the other textual data sources.

Data gathering also plays an important role. Although there are some studies that extract beyond direct firm-related data \cite{n-12}, \cite{n-194}, vast majority of the existing studies focus solely on the direct company-specific texts. Using sector, market related data, extending the sphere of firm-related texts (e.g. including not just stock related, but also product, service related texts as well) can be further developed and explored for stock market analysis.

\bibliographystyle{ACM-Reference-Format}
\bibliography{main}

\end{document}